%% file: n.tex
\documentclass[12pt]{article}
\usepackage{epsfig}

\input econfmacros.tex
\def\Title#1{\begin{center} {\Large {\bf #1} } \end{center}}

\def\Eej{E_{\rm ej}}

\def\tobs{t_{\rm obs}}

\def\sT{\sigma_{\rm T}}

\def\mdec{m_{\rm dec}}
\def\Rdec{R_{\rm dec}}
\def\Rgap{R_{\rm gap}}
\def\Racc{R_{\rm acc}}
\def\macc{m_{\rm acc}}

\def\Rload{R_{\rm load}}

\def\tpeak{t_{\rm flash}}
\def\Eflash{E_{\rm flash}}
\def\Gsh{\Gamma_{\rm sh}}
\def\Gej{\Gamma_{\rm ej}}

\def\dd{\rm d}

\def\be{\beta}
\def\Gn{\Gamma_n}
\def\bn{\beta_n}
\def\Gej{\Gamma_{\rm ej}}

\def\Gsh{\Gamma_{\rm sh}}

\def\Mej{M_{\rm ej}}
\def\tobs{t_{\rm obs}}
\def\Rdec{R_{\rm dec}}
\def\Rtr{R_{\rm trail}}

\def\mdec{m_{\rm dec}}
\def\be{\begin{equation}}
\def\ee{\end{equation}}
\def\sT{\sigma_{\rm T}}

\def\dM{\dot{M}}

\def\Rb{R_\beta}
\def\mb{m_\beta}
\def\taub{\tau_\beta}

\def\Eej{E_{\rm ej}}

\def\Racc{R_{\rm acc}}

\def\Grel{\Gamma_{\rm rel}}

\def\dM{\dot{M}}

\def\sT{\sigma_{\rm T}}
\def\xiacc{\xi_{\rm acc}}
\def\xiload{\xi_{\rm load}}
\def\xigap{\xi_{\rm gap}}
\def\facc{f_{\rm acc}}
\def\Racc{R_{\rm acc}}
\def\Rgap{R_{\rm gap}}
\def\Rload{R_{\rm load}}
\def\Gej{\Gamma_{\rm ej}}
\def\dd{{\rm d}}

\def\Eej{E_{\rm ej}}

\def\Eej{E_{\rm ej}}
\def\Grel{\Gamma_{\rm rel}}
\def\Rdec{R_{\rm dec}}
\def\mdec{m_{\rm dec}}
\def\tobs{t_{\rm obs}}

\newbox\grsign \setbox\grsign=\hbox{$>$} \newdimen\grdimen \grdimen=\ht\grsign
\newbox\simlessbox \newbox\simgreatbox \newbox\simpropbox
\setbox\simgreatbox=\hbox{\raise.5ex\hbox{$>$}\llap
     {\lower.5ex\hbox{$\sim$}}}\ht1=\grdimen\dp1=0pt
\setbox\simlessbox=\hbox{\raise.5ex\hbox{$<$}\llap
     {\lower.5ex\hbox{$\sim$}}}\ht2=\grdimen\dp2=0pt
\setbox\simpropbox=\hbox{\raise.5ex\hbox{$\propto$}\llap
     {\lower.5ex\hbox{$\sim$}}}\ht2=\grdimen\dp2=0pt
\def\simgt{\mathrel{\copy\simgreatbox}}
\def\simlt{\mathrel{\copy\simlessbox}}

\begin{document}

\Title{Interaction of GRB fireballs with ambient medium}

\bigskip\bigskip

%+\addtocontents{toc}{{\it D. Reggiano}}
%+\label{ReggianoStart}

\begin{raggedright}  

{\it Andrei M. Beloborodov\index{Beloborodov, A.M.}\\
Canadian Institute for Theoretical Astrophysics\\
University of Toronto, 60 St. George Street \\
Toronto, M5S 3H8 Ontario, CANADA}
\bigskip\bigskip
\end{raggedright}

%\begin{abstract}

%\end{abstract}

\section*{Abstract}

The customary picture of the fireball deceleration by an external medium
neglects two physical agents in the problem that are important at
radii $R<10^{17}$~cm: a radiative precursor (the prompt GRB) and a 
relativistic front of free neutrons. 
The radiative precursor preaccelerates the medium and loads it with 
$e^\pm$ pairs. This provides a new explanation for the early optical
flashes in GRBs. The front of free neutrons must form in a standard baryonic
fireball and change the mechanism of the fireball deceleration.
The neutron effect, however, disappears if the fireball is strongly 
dominated by the Poynting flux. The neutrons thus provide a unique link 
between the progenitor physics and the observed external blast wave.
The explosion mechanism at $R<10^{17}$~cm is testable with 
upcoming observations of early afterglows.

\section{Introduction}

The picture of an ultra-relativistic shell (``fireball'') ejected by a 
central engine became a customary model of GRBs. The shell emits the burst 
of $\gamma$-rays, sweeps up a static ambient medium, and decelerates, 
producing the afterglow emission. This simple model is consistent with 
the observational data. Recently, however, it has been realized to be 
intrinsically inconsistent at radii up to $R\sim 10^{17}$~cm. 
The picture is missing two physical agents that appear inevitably in 
the model: 

(1) The prompt burst of $\gamma$-rays acts as a powerful precursor ahead
of the fireball. It dramatically changes the early phase explosion, at radii
$R<2\times 10^{16}E_{53}^{1/2}$~cm where $E$ [erg] is the isotropic energy 
equivalent of the $\gamma$-ray burst.

(2) The standard GRB fireball contains free neutrons. An interesting 
fraction of neutrons survive up to $R\approx 8\times 10^{16}(\Gej/300)$~cm 
and changes the mechanism of the fireball-medium interaction.

Here, we describe these advances in the fireball physics.
The radiative precursor turns out to have an important connection with the 
prompt optical flashes observed in GRBs --- it offers a new explanation for 
the flash, alternative to the reverse-shock interpretation; this explanation 
works for the Poynting-dominated as well as barynic fireballs.

%##########################################################################

\section{Radiative precursor}

Radiative precursors are known to be important for interstellar
shocks~\cite{Draine}. 
In the case of GRB explosion, there is an evident fierce precursor: the 
initial burst of $\gamma$-rays with the isotropic energy equivalent up
to $3\times 10^{54}$~erg. The $\gamma$-rays come to the observer before the 
afterglow,\footnote{In fact, the $\gamma$-rays can be emitted quite early
inside the fireball, preceding the development of the external blast wave. 
The strong ms variability in GRB light curves is often cited as an evidence 
for small radius of $\gamma$-ray emission $R\sim 10^{12}-10^{14}$~cm; 
the ensuing blast wave radiates at $R\sim 10^{15}-10^{17}$~cm. 
}
so they must propagate ahead of the blast wave and impact the external medium.
This loophole of the GRB model has been noted recently~\cite{Mad1}-\cite{Bel1}. 
Two major effects of the $\gamma$-ray precursor are preacceleration of 
the ambient medium and its loading with $e^\pm$ pairs. 
The exact numerical calculation of the medium dynamics in the 
precursor has been done and understood analytically~\cite{Bel1}. 
It is summarized and explained below.

Let $dE/dS$ be the energy column density of the $\gamma$-ray front 
[erg/cm$^2$].  The GRB is likely beamed, yet it is convenient to define its
isotropic equivalent $E=4\pi R^2(dE/dS)$. When an ambient electron
is overtaken by the radiation front, it scatters energy 
$e_{\rm sc}=\sT(dE/dS)=E\sT/4\pi R^2$ (with a slight modification due to
Klein-Nishina correction to Thomson cross section $\sT$). The $e_{\rm sc}$ 
should be compared with the electron rest-mass energy, and this gives the
relevant dimensionless parameter,
\begin{equation}
\label{eq:xi}
  \xi=\frac{e_{\rm sc}}{m_ec^2}=65E_{53}R_{16}^{-2}.
\end{equation}
One would like to know the velocity $\beta$ acquired by the medium behind 
the radiation front and the number of loaded pairs per one ambient electron, 
$f=n_\pm/n_e$. The medium dynamics in the front is nonlinear because
the scattered photons turn into $e^\pm$ pairs which
do more scattering, and the problem is further
complicated by the transfer of the scattered radiation across the front.
Yet an exact solution can be obtained and well approximated
by a simple analytical model~\cite{Bel1},
\begin{equation}
  \label{eq:gam}
   \gamma(\xi)=
  \left\{\begin{array}{ll}
    1 & \xi<\xiacc, \\
    (\xi/\xiacc)^3 & \xiacc<\xi<3\xiacc, \\
  3\sqrt{3}(\xi/\xiacc)^{3/2} & \;\; \xi>3\xiacc,\\
  \end{array}\right.
\end{equation}
\begin{equation}
\label{eq:f}
   f(\xi)=
  \left\{\begin{array}{ll}
    \frac{1}{2}[\exp(\xi/\xiload)+\exp(-\xi/\xiload)] & \xi<\xiacc, \\
    (\xi/\xiacc)^2\facc & \xiacc<\xi<3\xiacc, \\
  3(\xi/\xiacc)\facc & \xi>3\xiacc,\\
  \end{array}\right.
\end{equation}
where $\xiload=20-30$, depending on the spectrum of the gamma-rays,
$\xiacc=5\xiload=100-150$, and 
$\facc=[\exp(\xiacc/\xiload)+\exp(-\xiacc/\xiload)]/2=74$.

If $\xi<\xiload$ nothing interesting happens with the medium in the 
$\gamma$-ray front: it remains almost static and $e^\pm$-free. If the front has 
$\xi>\xiload$, the runaway $e^\pm$ loading occurs. The number of loaded pairs 
depends exponentially on $\xi$ as long as $\xi<\xiacc$.
The front acts as a relativistic accelerator if $\xi>\xiacc$,
and the medium Lorentz factor $\gamma$ grows with $\xi$ as a power-law.
At $\xi=\xigap\approx 3\times 10^3$, 
$\gamma$ exceeds the Lorentz factor of the ejecta; it implies that the 
radiative precursor pushes the external medium away from the fireball and 
opens a gap.

The $\gamma$-ray front expands with time, and its 
$\xi$-parameter evolves as $\xi\propto R^{-2}$ (Eq.~\ref{eq:xi}), i.e.,
decreases.
It starts at very high $\xi$ and then passes through $\xigap$, $\xiacc$,
and $\xiload$ at radii $\Rgap$, $\Racc$, and $\Rload$, respectively, 
\be
\label{eq:Racc}
   \Racc\approx 10^{16} E_{53}^{1/2}{\rm ~cm}, \qquad \Rgap\approx \Racc/3,
   \qquad \Rload=2.3\Racc.
\ee
The three characteristic radii define four stages of the explosion:
\medskip

{\bf I.} $R<\Rgap$. The ejecta move in a cavity cleared by
the radiation front. The ambient medium surfs ahead with $\gamma>\Gej$.

{\bf II.} $\Rgap<R<\Racc$. The ejecta sweep the $e^\pm$-rich medium that
has been preaccelerated to $1\ll\gamma<\Gej$.

{\bf III.} $\Racc<R<\Rload$. The ejecta sweep the 
practically static medium ($\gamma\approx 1$). The medium is, however,
dominated by the loaded $e^\pm$.

{\bf IV.} $R>\Rload$. The ejecta sweep the static pair-free medium.
The precursor effect is negligible.
\medskip

\noindent
The blast wave develops at $R>\Rgap$. Technically, its model is constructed 
in the same way as it was done for a static external medium.
The only complication is that now the preshock material is $e^\pm$-rich and
moving relativistically. At $R=\Rgap$ the blast wave
gently begins to sweep up the preaccelerated medium with a small relative
Lorentz factor, $\Grel\approx\Gej/\gamma\approx 1$. With increasing $R>\Rgap$,
$\gamma$ falls off quickly and approaches $\gamma=1$ at $R=\Racc$ as
$\gamma=(R/\Racc)^{-6}$. Thus, the fireball suddenly ``learns'' that there 
is an interesting amount of relatively slow material on its way and hits it 
with a large $\Grel$. This resembles a collision with a wall and
results in a sharp rise in dissipation at $R\approx\Racc$
(see Fig.~4 in~\cite{Bel1}).
This dynamic effect is especially dramatic for GRBs with massive 
progenitors. Their ambient medium is a dense wind from the progenitor
and the swept-up mass at $\Racc$, $\macc=m(\Racc)$, (the ``wall'') can
exceed $\Eej/\Gej^2c^2$, where $\Eej$ is the total fireball energy.
Then the delayed collision with $\macc$ is the main dissipation event 
in the history of the explosion.

We note here that the optically thin medium scatters only a small portion 
of the $\gamma$-ray energy. In view of this fact, 
the strong dynamical impact of the $\gamma$-ray precursor on the fireball
deceleration might seem surprising. The clue to this paradox is the high 
Lorentz factor of the fireball, $\Gej\gg 1$. Because of high $\Gej$, 
a small ambient mass is expected to efficiently decelerate the fireball: 
half of $\Eej$ would be dissipated when the fireball sweeps up a static 
$m=\Eej/\Gej^2c^2$~\cite{Rees}. 
If, however, the fireball sends ahead a radiation 
precursor, it easily preaccelerates the medium to $\gamma\gg 1$ by 
depositing a relatively small energy $\gamma mc^2\ll\Eej$. Even at 
$\gamma>\Gej$, the deposited energy is small, while the dynamical impact is 
enormous: $m$ runs away and the fireball moves freely in the cavity cleared
by the $\gamma$-ray precursor. Thus, the fireball dissipation is delayed 
until the precursor is diluted sufficiently by side expansion.

The described scenario is modified if the $\gamma$-ray front is 
created by the blast wave itself rather than by internal dissipation 
in the fireball. Then a self-consistent blast wave model can be obtained
without gap opening. If, however, the ambient medium is sufficiently dense
then an ``electro-magnetic catastrophe'' can occur in the forward shock
as described in~\cite{Stern}. The runway production of $\gamma$-rays 
may result in a temporary gap openning and complicated temporal evolution 
of the shock. The feedback of the $\gamma$-ray precursor on the 
shock is a major agent governing this evolution.

%###########################################################################

\section{The prompt optical flash}

The optical flash was first detected in GRB~990123~\cite{Aker}.
It occurs much earlier than the normal optical 
afterglow that is observed days after the $\gamma$-ray burst.
In GRB~990123, the optical flash peaked less than minute after the 
GRB trigger. It was 
explained (and in fact predicted) as a possible emission from the reverse 
shock in the ejecta~\cite{Mesz2,Sari}. 
This explanation would exclude the Poynting-flux models of GRB ejecta where 
the reverse shock is energetically unimportant or does not occur at all. 
This explanation, however, is not unique. 

In fact, even in the absence of any emission from the reverse
shock, one expects a soft flash from the {\it forward} shock~\cite{Bel1}.
The blast wave evolution described above implies that the afterglow starts 
with a brief and very soft flash. It results from 
the pair loading and preacceleration of the ambient medium by the 
$\gamma$-ray front.

One can evaluate the effect using the 
customary synchrotron emission model. In the blast frame, the synchrotron 
spectrum of $e^\pm$ with Lorentz factor $\gamma_e$ peaks at  
\be
  \tilde{\nu}_s=10^6 B\gamma_e^2{\rm ~Hz},
\ee
where $B$ is the postshock magnetic field. The observed peak frequency 
is $\nu_s\approx\Gamma\tilde{\nu}_s$, where $\Gamma$ is the current 
Lorentz factor of the blast. $B$ is usually parametrized
in terms of the equipartition value, so that $B^2/8\pi$ is a fraction 
$\epsilon_B$ of the total energy density. Using the 
jump conditions at the shock front, one can write 
\begin{equation}
 B=c\frac{\Gamma}{\gamma}\sqrt{\frac{32\pi\epsilon_B\rho_0}{\gamma(1-\beta)}}.
\end{equation}
Thus, preacceleration of the preshock medium to $\gamma>1$ reduces $B$ by 
a factor of $\gamma^{-1/2}$. 

A fraction $\epsilon_e\sim 0.01-0.1$ of the energy of shocked ions is 
shared by $e^\pm$. The mean randomized Lorentz factor of the postshock 
$e^\pm$ is 
\be
  \gamma_e\approx\frac{\Gamma}{\gamma}\frac{m_p}{fm_e}\epsilon_e,
\ee
where $f=n_\pm/n_e$ is the pair loading factor.
Substituting Eqs.~(6) and (7) into (5), one finds that the $\gamma$-ray 
precursor reduces the synchrotron peak frequency $\nu_s$ by the factor of 
$\gamma^{-5/2}f^{-2}$. This is a big effect at all 
$R<\Rload\approx 2\times 10^{16}E_{53}^{1/2}$~cm. The afterglow should
start as a very soft signal, and it quickly hardens as $\gamma(R)$ and $f(R)$ 
decrease steeply with radius [see Eqs.~(1-3)]. 
A maximum $\nu_s$ is achieved at $\Racc\simlt R\simlt\Rload$, and then
the usual decay sets in (see also Fig.~6 in~\cite{Bel1}). 
 
The initial soft flash should have a broad spectrum and can be observed in 
the optical band. It appears to be an inevitable result of the 
preacceleration and $e^\pm$-loading of the ambient medium.
Because the flash is emitted by the forward shock, this mechanism works 
regardless the nature of the GRB ejecta. It applies to Poynting-flux ejecta 
as well as to standard baryonic fireballs. 

One can easily evaluate the arrival time of the expected flash.
The flash should peak at observed time
\begin{equation}
\label{eq:tpeak}
  \tpeak\approx\frac{\Racc}{2\Gej^2}(1+z) 
 \approx 12\,E_{53}^{1/2}\left(\frac{\Gej}{100}\right)^{-2}
  \left(\frac{1+z}{3}\right) {\rm~s},
\end{equation}
where $z$ is the cosmological redshift of the burst.
A nice feature of this prediction is that $\tpeak$ 
depends only on the isotropic equivalent of 
the burst energy, $E$, and the initial Lorentz factor of the ejecta, $\Gej$.
If one manages to observe the peak of the flash
and measure the burst redshift (which gives $E$), then one can infer 
$\Gej$ --- the unknown and probably most important parameter of the 
GRB phenomenon. 
The optical flash in GRB~990123 peaked at $\tpeak\approx 20$~s 
(corrected for cosmological time dilation)
and the isotropic $\gamma$-ray energy of the burst was 
$E\approx 3\times 10^{54}$~erg~[7]. Eq.~(\ref{eq:tpeak}) then implies
$\Gej\approx 200$. This result does not depend on $\epsilon_B$, $\epsilon_e$,
$n_0$, the beaming angle and the nature of the ejecta, or any other 
%poorly known 
parameters of the explosion.

The energy of the pair-loaded flash is approximately
\be
  \Eflash\sim \macc c^2\Gamma^2,
\ee
where $\macc$ is the ambient mass within $\Racc$. 
If $\macc<\Eej/\Gej^2c^2$ (i.e. $\Racc$ is smaller than the 
characteristic deceleration radius of the fireball $\Rdec$) then 
$\Gamma\approx \Gej$ during the flash; the main peak of the afterglow 
occurs later at $R=\Rdec$ and emits more energy (in the X-ray bands). 
Then $\Eflash$ is a small fraction of the total energy of the explosion.
In the case of a homogeneous ambient medium with 
density $n_0=const$, one finds~\cite{Bel1}
\be
   \frac{\Eflash}{\Eej}\approx 10^{-3}n_0
       \left(\frac{\Gej}{300}\right)^2 \frac{E_{53}^{3/2}}{{\Eej}_{53}}.
\ee

GRBs with massive progenitors occur inside dense 
star-forming clouds with $n_0\sim 1-10^2$~cm$^{-3}$ and
a detectable flash is expected. Even more importantly, the progenitor 
itself emits a powerful wind all the time before it explodes, and its
wind is the actual ambient medium of the GRB (with density scaling 
with radius as $R^{-2}$ out to a parsec distance).
%~\cite{Li}. 
The wind from a Wolf-Rayet progenitor can easily have $\macc>\Eej/\Gej^2c^2$.
Then the main fireball dissipation occurs at $R\sim \Racc$ in the medium
heavily loaded with $e^\pm$; the resulting powerful soft flash coincides 
with the main peak of the afterglow~\cite{Bel1}.

By now the early optical emission has been detected in four bursts:
GRB~990123, GRB~021004, GRB~021211, and GRB~030329. The initial peak  
of the optical emission has been seen only in GRB~990123.
This is naturally explained by the model described above: 
the flash should peak earlier than 10~s unless 
$\Gej<100E_{53}^{1/2}(1+z)^{1/2}$ (see Eq.~\ref{eq:tpeak}), and therefore 
it is difficult to detect the peak with current instruments. The proposed 
microsatellite {\em ECLAIRs}~\cite{ECLAIRs} would be very helpful as it 
can provide data in soft bands at all times, even much shorter than 10~s.

%##########################################################################

\section{Neutron front}

An intriguing aspect of the GRB explosion, again related to the 
ultra-relativistic character of the explosion, is a significant 
contamination of the fireball by free neutrons~\cite{Der1}-\cite{Bel2}. 
%(Derishev, Kocharovsky, \& Kocharovsky 1999a,b; Bahcall \& M\'esz\'aros 
%2000; M\'esz\'aros \& Rees 2000; Fuller, Pruet, \& Abazajian 2000; 
%Beloborodov 2003a,b). 
Besides opening a possibility to observe multi-GeV neutrinos generated by 
neutron-ion collisions, it has important implications for the afterglow 
physics: the very mechanism of the afterglow can be changed by neutrons 
at radii up to $10^{17}$~cm~\cite{Bel3}.

\subsection{Neutronization of the Central Engine}

What is ejected in a GRB explosion?
Its engine is very dense and hot, and 
the central matter (or the matter surrounding the central black hole) is 
filled with $e^\pm$ pairs in thermodynamic equilibrium with blackbody 
radiation at a temperature $kT=1-10$~MeV. Its baryonic component is made
of neutrons and protons (composite nuclei break up into free nucleons in 
the unshadowed region of the $T-\rho$ plane in Fig.~1) 
and frequent $e^\pm$ captures onto nucleons take place,
\begin{equation}
  e^-+p\rightarrow n+\nu, \qquad e^++n\rightarrow p+\bar{\nu}.
\end{equation}
These reactions quickly convert protons to neutrons and back, and establish 
an equilibrium proton fraction $Y_e=n_p/(n_n+n_p)$ which is shown in Figure~1. 
Any plausible GRB engines belong to the $T-\rho$ region where $Y_e<0.5$
(see \cite{Bel2} and refs. therein). So, the neutrons are not only present 
in the central engines --- they dominate.

%%%%%%%%%%%%%%%%%%%%%%%%%%%%%%%%%%%%%%%%%%%%%%%%%%%%%%%%%%%%%%%%%%%%%%%%%
\begin{figure}[htb]
\begin{center}
\epsfig{file=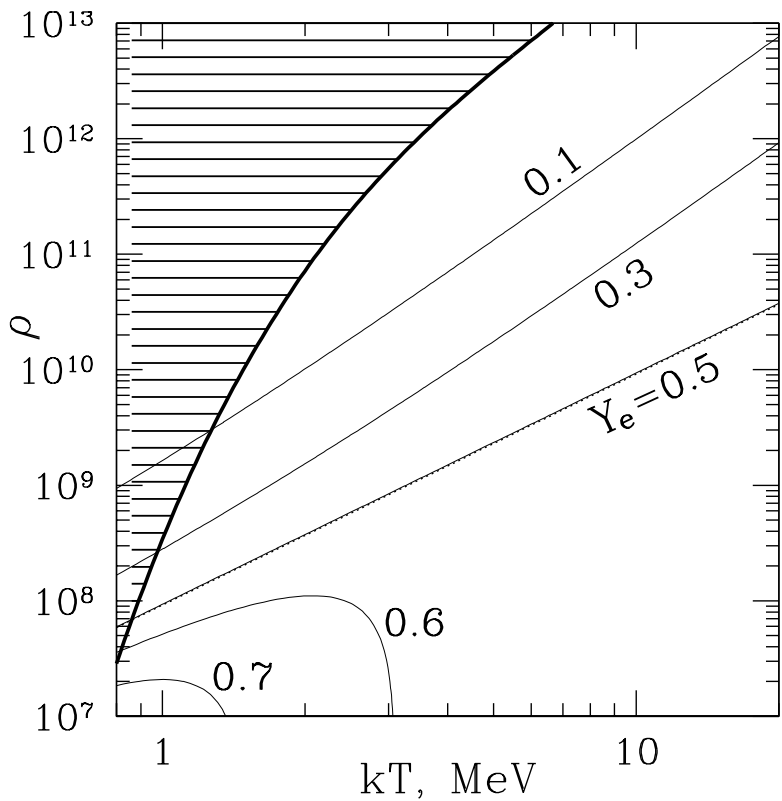,height=10.5cm}
\caption{Contours of equilibrium $Y_e=n_p/(n_n+n_p)$ on
the $T$-$\rho$ plane for $\nu$-transparent
matter. Thick curve shows the boundary of free-nucleon region.
A similar plot for $\nu$-opaque matter is shown in~\cite{Bel2}.
}
\label{fig:nuc}
\end{center}
\end{figure}
%%%%%%%%%%%%%%%%%%%%%%%%%%%%%%%%%%%%%%%%%%%%%%%%%%%%%%%%%%%%%%%%%%%%%%%%%%%

When the neutronized matter is ejected into a high-entropy fireball, it 
remains neutron-dominated because (1) expansion occurs too fast 
and $Y_e$ freezes out at its initial value below 0.5, or (2) the fireball 
partially absorbs the neutrino flux from the engine, which tends to keep 
$Y_e<0.5$ (see \cite{Bel2} for details).

%#######################################################################

\subsection{Ejected Neutrons Survive the Nucleosynthesis}

As the fireball expands and cools, free nucleons tend to
recombine into $\alpha$-particles (like they do in the big bang). 
This process competes, however, with rapid expansion and can freeze out. 
For this reason, nucleosynthesis is suppressed in fireballs with a high 
photon-to-baryon ratio $\phi=n_\gamma/n_b$ (or, equivalently, high entropy 
per baryon $s/k=3.6\phi$). A minimum $\phi$ in GRBs is $\sim 10^5$, which is
just marginal for nucleosynthesis (detailed nucleosynthesis calculations
have been done recently in \cite{Bel2},\cite{Lem},\cite{Pru1}).
Even in the extreme case of complete helium production, however, there 
are still leftover neutrons because of the neutron excess ($Y_e<0.5$).
The nucleon recombination into $\alpha$-particles consumes equal numbers
of $p$ and $n$, and the minimum mass fraction of leftover neutrons is 
$X_n=1-2Y_e$.\footnote{For a similar reason, 75\% of baryons in the universe
are protons. There was a proton excess at the time of primordial 
nucleosynthesis with $Y_e\approx 7/8$. After nucleon recombination, 
it resulted in mass fraction $X_p=2Y_e-1=0.75$ of leftover protons.}
%The minimum mass fraction of neutrons is $X_n=1-2Y_e$.

In addition, the synthesized helium is likely destroyed during the ensuing 
evolution of the explosion~\cite{Bel2}. 
This happens if (1) there appears a significant 
relative bulk velocity between the neutron and ion components (as a result 
of neutron decoupling during the acceleration stage of the fireball) or 
(2) internal shocks occur and heat the ions to a high temperature. 

Thus, a substantial neutron component appears inevitably in the standard 
fireball scenario. The neutrons develop a Lorentz factor $\Gn=10^2-10^3$ 
at the very beginning of the explosion when the fireball is accelerated 
by radiation pressure; they are collisionally coupled to the ions in the
early dense fireball, and decouple close to the end of the acceleration 
stage. Then the neutrons coast and gradually decay with a mean lifetime
$\taub\approx 900$~s and a mean decay radius $\Rb=c\taub\Gn$,
\be
\label{eq:H}
 \Rb=0.8\times 10^{16}\left(\frac{\Gn}{300}\right) {\rm ~cm}.
\ee
We will show below that neutrons change the fireball-medium interaction
at radii as large as $10\Rb$.

%#######################################################################

\subsection{Neutron-Fed Blast Wave}

Let us remind what happens in a relativistic explosion without neutrons. 
The ejected fireball with mass $\Mej=\Eej/\Gej c^2$ and Lorentz factor 
$\Gej$ sweeps up 
an ambient medium with density $n_0=1-10^2$~cm$^{-3}$ and gradually dissipates 
its kinetic energy. The dissipation rate peaks at a characteristic 
``deceleration'' radius 
$\Rdec\sim 10^{16}$~cm where half of the initial energy is dissipated.
$\Rdec$ corresponds to swept-up mass $\mdec=\Mej/\Gej$. Further
dynamics is described by the self-similar blast wave model.
How does this picture change in the presence of neutrons?

At radii under consideration, $R>10^{15}$~cm, the ejected fireball is a
shell of thickness $\Delta\ll R$. In contrast to neutrons, the ion component
of the fireball is aware of the external medium and its Lorentz factor
$\Gamma$ decreases. As $\Gamma$ decreases below $\Gamma_n$, the ions fall
behind and separate from the neutrons. Thus the fireball splits into two
relativistic shells which we name N (neutrons) and I (ions).
The mass of the leading N-shell is decreasing because of the $\beta$-decay,
\be
\label{eq:Mn}
  M_n(R)=M_n^0\exp\left(-\frac{R}{\Rb}\right).
\ee
The N-shell energy, $E_n=\Gn M_nc^2$, remains, however, huge compared to 
the ambient rest-mass $mc^2$ even at $R>\Rb$. For example, at $R=\Rdec$ one
finds $E_n/\mdec c^2=X_n\Gn\Gej\exp(-\Rdec/\Rb)$ where $X_n=M_n^0/\Mej$ is 
the initial neutron fraction of the fireball.

The neutron decay products $p$ and $e^-$ share immediately their huge
momentum with ambient particles due to two-stream instability
and the  N-shell leaves behind a mixed trail with a relativistic
bulk velocity $\beta<\bn=(1-1/\Gn^2)^{1/2}$~\cite{Bel3},
\be
\label{eq:beta}
  \beta(R)=\frac{\bn}{1+(\Gn\zeta)^{-1}},  \qquad
  \gamma(R)=\frac{1}{(1-\beta^2)^{1/2}}
           =\frac{\Gn\zeta+1}{(\zeta^2+2\Gn\zeta+1)^{1/2}},
\ee
where 
\be
\label{eq:zet}
  \zeta(R)=\frac{\dd M_n}{\dd m}
              =\frac{M_n}{\Rb}\left(\frac{\dd m}{\dd R}\right)^{-1},
\ee
and $m(R)$ is ambient mass enclosed by radius $R$. There exists
a characteristic radius $\Rtr$ where the trail becomes nonrelativistic
($\beta=0.5$). It is defined by condition $\zeta=\Gn^{-1}$ [see Eq.~(14)],
which requires about 10 e-foldings of the decay
(for a typical $\mb\sim\mdec\sim 10^{-5} M_n^0$). Thus,
\be
 \Rtr\approx 10\Rb=0.8\times 10^{17}\left(\frac{\Gn}{300}\right){\rm ~cm}.
\ee
$\Rtr$ depends very weakly (logarithmically) on the ambient density and the
initial neutron fraction of the fireball, $X_n$.

The decaying N-shell not only accelerates the ambient medium. 
It also compresses the medium, loads with new particles, and heats 
to a high temperature. The rest-frame density and relativistic 
enthalpy of the trail are
\be
\label{eq:mu}
  n=n_0(1+\zeta)\left(\zeta^2+2\Gn\zeta+1\right)^{1/2}, \qquad
  \mu=\frac{(\zeta^2+2\Gn\zeta+1)^{1/2}}{1+\zeta},
\ee
where $n_0$ is the medium density ahead of the N-shell. 
For $\Gn^{-1}<\zeta<\Gn$ one finds $\mu\gg 1$, i.e. 
the thermal energy of the trail far exceeds its rest-mass energy.

The ion fireball follows the neutron front and collects the trail.
As a result, (1) the ion Lorentz factor $\Gamma$ decreases and (2) a shock
wave propagates in the trail material. The shock has a Lorentz factor 
$\Gsh\simgt\Gamma$ and it cannot catch up with the neutron 
front (unless $n_0[R]$ falls off steeper than $R^{-3}$). Dynamics and 
dissipation in the shock are discussed in \cite{Bel3}. We emphasize here 
that the shock propagates in a relativistically moving, dense, hot, 
and possibly magnetized medium left behind the leading neutron front. 
It is very different from the customary shock in a cold and static 
interstellar medium. A similar situation may take place for fireballs 
expanding in plerionic surroundings~\cite{Kon}.

The neutron impact ceases at $\Rtr\approx 10^{17}$~cm, which can leave an
imprint on the observed afterglow. For example, the shock dissipation can
have a second bump~\cite{Bel3}, and a spectral transition is also 
possible. The arrival time of radiation emitted at $\Rtr$ is approximately
$\Rtr/2\Gamma^2 c$ (counted from the arrival of first
$\gamma$-rays). It may be as long as 30 days or as short as a few seconds,
depending on the fireball Lorentz factor $\Gamma(\Rtr)$. 
Recent early observation of a GRB afterglow (GRB~021004) discovered a
strong re-brightening at $10^3$~s, which may be related to neutrons.
Also, we do not exclude a possible relevance of neutrons to the 20~day 
bumps observed in a few GRBs, as the time coincides with $\Rtr/c$.

Neutron signatures should be absent if the fireball is 
dominated by a Poynting flux and has extremely low baryon loading. Then the 
neutron component decouples early, with a modest Lorentz factor $\Gn$, and 
decays at small radii, $R\ll\Rdec$. The upper bound on $\Gn$ due to 
decoupling is $\Gn\approx 300(\dM_\Omega/10^{26})^{1/3}$ where 
$\dM_\Omega$ [g/s] is the 
mass outflow rate per unit solid angle of the fireball~\cite{Bel2}.

\section{Conclusions and prospects}

The deceleration radius of GRB fireballs is believed to be in the range
$10^{15}-10^{17}$~cm, depending on the ambient density and the energy
of the explosion. It falls in the interesting range where the prompt 
$\gamma$-rays and free neutrons strongly impact the fireball interaction
with the external medium. Therefore, the blast wave, at least at its 
early stage, should involve the described effects. The upcoming 
data from {\it Swift} may be crucial for understanding the basic
mechanism of the fireball deceleration. 
Recent observations of variable early afterglows in GRB~021004 and GRB~030329 
indicate that additional complications may be involved such as clumpy ambient 
medium or refreshed afterglows.

The explosion scenario and its observational signatures depend on the nature 
of the ejecta.
If the fireball is dominated by Poynting flux, the reverse shock emits little
energy, and the observed early optical flash must be dominated by the 
forward shock in the manner described in section~3. 
If the fireball is dominated by baryons rather than Poynting flux,
the reverse shock can also contribute to the flash.

The formation of a leading neutron front appears to be a robust feature of 
a baryonic fireball, and detection of neutron signatures in the afterglow
emission would be important.
The current poor understanding of the shock physics and particle acceleration
in GRBs impede, however, definite predictions for such signatures,
and the model needs to be developed in this direction.
For bright bursts, $E\sim 10^{54}$~erg, the pair-loading radius 
$\Rload$ of the $\gamma$-ray precursor gets comparable to $\Rtr$, and 
then the neutrons and $\gamma$-rays have a combined impact on the ambient
medium. This situation has not been studied yet. 

Afterglows observed to date typically have first data points 
hours or days after the prompt GRB. 
The main peak of the blast wave emission is missed. It is expected 
at seconds or minutes (and possibly overlaps with the prompt 
$\gamma$-rays). The observed tail of the afterglow comes from the 
slowed-down blast wave, which has an estimated radius $R\sim 10^{17}$~cm. 
With {\em Swift} it may be possible to observe routinely earlier afterglows 
that come from smaller radii. The afterglow is expected to 
%suddenly ``switch on'' and 
rise sharply in soft bands at observed time
$\tobs\sim 10E_{53}^{1/2}(\Gej/100)^{-2}(1+z)$~s.
The initial flash must be very soft, and it can emit significant energy before 
a normal X-ray afterglow sets in, especially in the massive progenitor 
scenario. The peak of this flash is difficult to detect because it happens 
quickly, however, future observations may overcome this technical difficulty. 
{\em Swift} should be able to see optical emission 30-70~s after the beginning 
of the burst, and the proposed {\em ECLAIRs} would detect flashes at much 
smaller times~\cite{ECLAIRs}. A detected initial flash of the afterglow would 
provide valuable information on the ejecta Lorentz factor, and its time 
profile and spectrum can give indications on the nature of the ambient medium, 
and hence the progenitor.

Additional diagnostics for the fireball-medium interaction will be provided 
by {\em GLAST}. At the afterglow radii, the ambient medium is likely 
optically thin (even after $e^\pm$ loading by the $\gamma$-ray precursor), 
and then the bulk of GRB radiation passes freely through it. 
However, the most energetic $\gamma$-rays may encounter a substantial 
$\gamma-\gamma$ opacity made by the scattered radiation.
If the ambient medium is a wind from a massive progenitor,
the prompt $\gamma$-rays should be absorbed above 
$\epsilon_{\rm br}=5-50$~MeV, depending on the density of the 
wind~\cite{Bel1}. This produces a spectral break that should be easily 
detected by {\em GLAST}. The scattered $\gamma$-rays may also be observed as 
an echo of the burst~\cite{Mad2}.

%##########################################################################

\def\Discussion{
\setlength{\parskip}{0.3cm}\setlength{\parindent}{0.0cm}
     \bigskip\bigskip      {\Large {\bf Discussion}} \bigskip}
\def\speaker#1{{\bf #1:}\ }
\def\endDiscussion{}

\end{document}

%% file: econfmacros.tex
\def\beq{\begin{equation}}
\def\eeq#1{\label{#1}\end{equation}}
\def\eeqn{\end{equation}}

\def\beqa{\begin{eqnarray}}
\def\eeqa#1{\label{#1}\end{eqnarray}}
\def\eeqan{\end{eqnarray}}

\let\bar=\overbar

\def\Dslash{\not{\hbox{\kern-4pt $D$}}}
\def\dslash{\not{\hbox{\kern-2pt $\del$}}}

\def\ee{e^+e^-}

\def\msb{{\bar{\ssstyle M \kern -1pt S}}}